\documentclass[11pt,a4]{article}
\usepackage{graphicx}
\usepackage{amsmath}
\usepackage[a4paper, total={6in, 8in}]{geometry}

\title{\bf Alloying Ratio Versus Cluster Size for Reversible Hydrogen Storage in Ni Doped Small Mg Clusters: Dispersion Corrected DFT Study}
\author{ Bishwajit Boruah, Bulumoni Kalita\footnote{Corresponding Author, bulumonikalita@dibru.ac.in}\\
Dept. of Physics, Dibrugarh University,Assam, India, 786001}

\begin{document}
\maketitle

\begin{abstract}
Dispersion corrected density functional theory ($\omega$B97X-D DFT) method is used to study the molecular hydrogen adsorption in $Ni_nMg_m$ $(1\geq n\geq 3,1\geq m\geq9)$ clusters.  All these clusters can effectively adsorb multiple $H_2$ in the preferred binding energy (BE) range between physisorption and chemisorption, i.e., $0.1 eV\geq BE \geq0.8 $eV. $H_2$ adsorption on $Ni_kMg_k$ (Ni:Mg=1:1), $Ni_kMg_{2k}$ (Ni:Mg=1:2) and $Ni_kMg_{3k}$ (Ni:Mg=1:3) (k=1-3) clusters shows fascinating behaviours in terms of Ni:Mg alloying ratio and cluster size.  In each Ni:Mg ratio, the number of adsorbed $H_2$ in the heavier clusters (k=2, 3) becomes integral multiples of that in the lightest configuration (k=1).  As a consequence, the gravimetric density of molecular hydrogen remains fixed at each Ni:Mg ratio irrespective of the cluster size.  The corresponding values are 17.94 wt\% (1:1), 14.46 wt\% (1:2) and 13.28 wt\% (1:3), which are significantly higher than the ultimate target of 6.5 wt\% set by DOE, US.  Molecular dynamics simulations further reveal that room temperature desorption of almost all $H_2$ molecules are possible for all the clusters. 
\end{abstract}

\section{Introduction}\label{sec1}

In the present scenario of growing energy demand, hydrogen has been is possibly emerging as the best alternative with its exceptionally impressive mass-energy density and cleanliness.  However, its safe storage targets set by the Department of Energy (DOE), USA   \cite{bib1} is still a challenge to accomplish.  The required binding energy of hydrogen with the storage material should be within the range of 0.1eV- 0.8 eV, i.e., in between physisorption and chemisorption processes \cite{bib1,bib2}. The optimum gravimetric density of hydrogen, which directly depends on the number of $H_2$ molecules adsorbed by storage material has to be 6.5 wt\% for commercial applications \cite {bib1}. In the recent years, interest has been given to the light elements such as Li, Be, B, C, Na, Mg, Al and K etc. due to their reasonably less molecular masses in order to achieve high gravimetric density of $H_2$ for practical applications \cite{bib4,bib5,bib6,bib7,bib8,bib9,bib10,bib11}.In a recent density functional theory (DFT) study, Jaiswal et al. have been able to achieve hydrogen gravimetric density of 18.7 wt\% in alkali metal decorated silicon clusters \cite{bib9}. In another study on aromatic bimetallic clusters, K. Srinivasu and co-workers found high hydrogen gravimetric density in hydrogenated $Al_4M_2, Be_3M_2, Mg_3M_2$ etc. (M=Li, Na, K) clusters.  They reported the formation of $Al_4M_2(H_2)_{2n}$ and $Be_3M_2(H_2)_{2n}$ complexes by $Al_4M_2$ and $Be_3M_2$ clusters on adsorption of multiple hydrogen molecules \cite {bib7}. Among the light elements, Mg based materials have been studied extensively in the recent years to design $H_2$ storage medium because of the advantageous properties such as cost effectiveness, abundance and non-toxicity etc \cite {bib12,bib13,bib14,bib15,bib16,bib17,bib18,bib19,bib20,bib21,bib22,bib23,bib24}.In a recent review, Xie et al. have discussed the first principles studies of particle size and dehydrogenation of $MgH_2$ and hydrogen adsorption properties of Mg surfaces, where it has been observed that Mg-H bond gets weakened with decreasing particle size \cite {bib12}. Yartys et al. have given a comprehensive report on the latest activities and historic overviews of Mg based hydrogen storage materials conferring the seriousness of improving their kinetics and thermodynamic properties \cite {bib15}. Theoretical and experimental studies reported that reducing the size of Mg particles to nanoscale can improve the kinetics and thermodynamic of hydrogen adsorption and desorption \cite {bib18,bib24,bib25,bib26,bib27}. Rudy et al. found that desorption energy of hydrogen decreases significantly as the crystal grain size goes down to 1.3 nm \cite {bib18}. Li et al. have reported that thinner Mg nanowires have lower desorption energy of hydrogen than the thicker Mg nanowires or bulk $MgH_2$\cite {bib24}. Jeon et al. reported that Mg nanoparticles can adsorb and release hydrogen below 200 ºC \cite {bib28}.DFT calculations have also shown that varying charge state, geometry, dopant element, alloying etc. one can tune the hydrogen adsorption properties of Mg clusters \cite {bib29,bib30}. Shen et al. have reported the saturated state for hydrogen chemisorption in Mg clusters to be $Mg_nH_{2n}$ \cite {bib29}. In another study Banerjee et al. noticed significant improvement on the hydrogen adsorption kinetic properties in Mg nanoclusters from their bulk counterpart \cite {bib30}. In another study, Srivastava et al. showed that hydrogen adsorption capacity of planar $(MgO)_n$ clusters increases linearly with the cluster size \cite {bib31}. Previous studies have shown that transition metal (TM) doping on Mg based materials can greatly influence their hydrogen adsorption behaviour \cite {bib32,bib33,bib34,bib35,bib36,bib37,bib38,bib39}. Trivedi and Bandopadhyay in their two different DFT studies revealed $Mg_5Co$ and $Mg_9Rh$ clusters to be very effective for hydrogen storage purpose \cite {bib34,bib36}.  Ma et al. have reported that adding Ti and Nb into $Mg_{55}$ cluster enhances the stability and also improve the hydrogenation kinetics \cite {bib38}.Very recently, Samantaray et al. have been able to synthesize NiMg alloy nanocomposites with up to 5.4 wt\% hydrogenation capacity \cite {bib40}. The issue of hydrogenation kinetics in $Mg_{17}M$ clusters doped with different 3d transition metal atoms (M) have recently appeared in a series of DFT studies by Charkin and Maltsev  \cite{bib41,bib42,bib43,bib44}. They have confirmed Mg17Ni cluster to have the most favourable reaction kinetics in forming $Mg_{17}MH_2$ systems  \cite{bib41}. 
\paragraph {}
From the above discussion, it becomes clear that nature of geometry and doping can play important roles in governing the hydrogen storage capacity of small clusters.  However, in spite of the number of exciting results, the effect of transition metal doping concentration on hydrogenation of Mg clusters has not been explored so far to the best of our knowledge.  Therefore, in the present study we aim to address this issue theoretically, i.e., we will carry out DFT studies to investigate systematically the effect of Ni doping concentration on the hydrogenation of small Mg clusters.

\section {Methodology}
The calculations are carried out in the Gaussian 09 package  \cite{bib45} using DFT method.  As Mg clusters possess van der Waals features and hence it is necessary to include dispersion correction in their calculations  \cite{bib46}. Moreover, dispersion correction is also necessary for accurate modelling of physisorption of molecular H2 on Mg clusters resulting from van der Waals interactions  \cite{bib30}.   In our recent study \cite{bib46},we have shown that dispersion corrected $\omega$B97X-D functional can precisely determine the structure and energetics of gas phase small $Mg_n$ clusters. The $\omega$B97X-D is a range-separated hybrid functional, which is capable of capturing both short-range and long-range interactions  \cite{bib47}. It is also reported that $\omega$B97X-D method is effective in the study of transition metal compounds \cite {bib48}. Therefore, we have used  $\omega$B97X-D functional for computing the gas phase $Ni_nMg_m$ $(1\geq n\geq 3,1\geq m\geq9)$ and their $H_2$ adsorbed complexes $((H_2)_xNi_nMg_m)$ in the present study.  6-311G(d,p) basis set is chosen for Mg and H atoms and LANL2DZ basis set is used for Ni atom.  We have used CALYPSO package to find the different possible structural isomers of NinMgm clusters \cite{bib49}. The binding energy per atom of the clusters are calculated using the following formula:

\begin{equation}
\label{simple_equation}
E_b= \frac{nE(Ni)+mE(Mg)-E(Ni_nMg_m)}{n+m}
\end{equation}
\\
The binding energy per $H_2$ molecule ($E_b$) is calculated using the following formula:
\begin{equation}
\label{simple_equation}
BE= \frac{E(Ni_nMg_m)+Ex(H_2)-E((H_2)_xNi_nMg_m)}{x}
\end{equation}
\\
Here n, m and x represent the number of Ni, Mg atoms and $H_2$ molecules, respectively.\\\\
The capacity of adsorption of multiple hydrogen molecules for the studied clusters is calculated in terms of wt\% as given by the following formula:
\begin{equation}
\label{simple_equation}
wt\%= \frac{x\mu(H_2)}{n\mu(Ni)+m\mu(Mg)+x\mu(H_2)}
\end{equation}
\\
Here $\mu$ represents the molecular mass of the corresponding atoms/molecules.
\section {Results and Discussions}
Prior to this work, we have studied the mutual effects of the electronic and geometric structures of neutral and cationic TMMg3 clusters on the nature of adsorption of single and multiple hydrogen molecules on these clusters \cite {bib50}.  We have found that a slightly higher energy less symmetrical planar isomer of $NiMg_3$ is capable of adsorbing larger number of $H_2$ molecules than the corresponding lowest energy symmetric ground state leading to the gravimetric density of 13.28 wt\%.  Therefore, in the present study, we have decided to consider only the planar like structures of $Ni_nMg_m$ clusters that are not necessarily the lowest energy ones.  Next, we are also interested in varying the alloying ratio of Ni:Mg in the same clusters and check their effect on molecular hydrogen storage.  Accordingly, we have chosen three Ni:Mg ratios viz. 1:1, 1:2 and 1:3 to investigate the hydrogen gravimetric densities of $Ni_nMg_m$ clusters.  Results for the three alloying ratios will be discussed separately in the following sections. 

\subsection {Geometry and Energetics}
\subsubsection {Hydrogen Adsorption on NinMgm clusters (n=m, n+m=2k, k=1-3)}
We have considered three different cluster compositions, NiMg, $Ni_2Mg_2$ and $Ni_3Mg_3$ in the 1:1 alloying ratio of Ni:Mg.  The initial geometries of all possible structural isomers of these clusters are first obtained running the CALYPSO code, which are then optimized in various spin multiplicities in Gaussian 09 package.  The resulted stable isomers are shown in Figure S1(a) in the supporting information (SI).  The most stable structures of the NiMg, $Ni_2Mg_2$ clusters resemble with those reported earlier \cite {bib51, bib52}.  We have checked the stability of the clusters from their binding energies calculated using equation (1) and from the absence of any imaginary vibrational frequency shown by the frequency calculations.  The binding energy value of NiMg dimer matches well with that in a previous report \cite {bib1}.  We have next studied the adsorption of multiple hydrogen molecules in all the stable clusters of NiMg, $Ni_2Mg_2$ and $Ni_3Mg_3$.  Interestingly, it has been observed that the maximum number of hydrogen molecules get adsorbed in some planar-like geometries of $Ni_2Mg_2$ and $Ni_3Mg_3$ clusters (Figure S1(b)).  These isomers with Ni atoms on the peripheral sites, however, possess relatively higher energy than the lowest energy isomers (Figure S1(a)).  Therefore, only these $Ni_nMg_m$ clusters capable of maximum $H_2$ adsorption and their saturated hydrogen-adsorbed complexes are shown in Figure 1(a) and 1(b), respectively.  Figure 1(c) shows the average Ni-Mg bond lengths of bare and hydrogenated clusters shown in Figure 1(a) and 1(b).  It is observed that the Ni-Mg bond length does not change much after hydrogenation and the cluster geometries are intact even after adsorption of multiple hydrogen molecules.  The slight decrement and increment in Ni-Mg bond lengths for $Ni_nMg_m$ clusters after hydrogenation is clearly seen in Figure 1(c).  The bond length of NiMg dimer matches well with the available results \cite {bib51, bib52}.  It is to be mentioned here that in view of preferred intermediate physisorption-chemisorption binding, the minimum cut-off value of binding energy per $H_2$ molecule has been considered to be -0.10 eV only for our study.  Accordingly, the NiMg dimer is found to adsorb up to 9 $H_2$ molecules within the optimum adsorption energy range of 0.10 eV – 0.80 eV.  Previously, adsorption of maximum of 7$H_2$ molecules was reported for $Ni_2$ dimer \cite {bib52}.  Therefore, our computed $(H_2)_9NiMg$ complex yields a reasonably higher hydrogen gravimetric density of 17.94 wt\%.  The hydrogen adsorption behaviour of the $Ni_2Mg_2$ and $Ni_3Mg_3$ clusters are quite surprising as they adsorb maximum of 18 and 27$H_2$ molecules, respectively, which are exactly the integral multiple of that number in NiMg dimer.  As a result, the gravimetric density of hydrogen in all of these clusters remains same.  On the basis of our results, we observe that linear and planar-like structures of $Ni_nMg_m$ clusters having 1:1 alloying ratio with sufficiently exposed Ni atoms can efficiently adsorb multiple numbers of hydrogen molecules.
\begin{figure}
\centering
\includegraphics[width=0.8\textwidth]{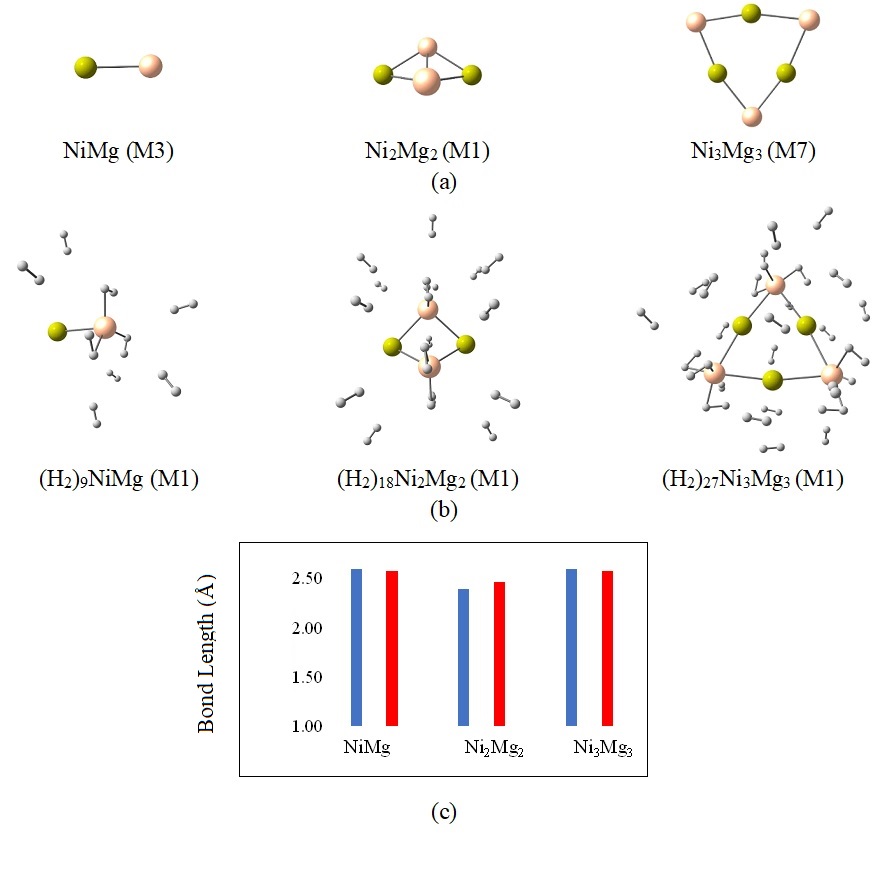}
\caption{\bf Geometries of (a) bare $\bf Ni_nMg_m$ clusters and (b) $\bf (H_2)_xNi_nMg_m$ complexes in 1:1 (Ni:Mg) ratio, (c) Ni-Mg bond lengths of bare and hydrogenated $\bf Ni_nMg_m$ clusters. (Colour blue and red are used to represent bare and hydrogenated complexes respectively, M represents spin multiplicity)}\label{Fig1}
\end{figure}
\newpage

\subsubsection {Hydrogen Adsorption on $Ni_nMg_m$ clusters (2n=m, n+m=3k, k=1-3)}
In the 1:2 (Ni:Mg) ratio of $Ni_nMg_m$ clusters, we have considered three compositions viz. $NiMg_2, Ni_2Mg_4$ and $Ni_3Mg_6$.  All the geometrical and spin isomers of these clusters are shown in Figure S2(a).  In case of these clusters also, maximum hydrogen adsorption is found in some planar-like isomers with the Ni atoms on the boundary positions (Figure S2(b)).  The corresponding structures of bare and hydrogen-adsorbed complexes are shown in figure 2(a) and 2(b).  The Ni-Mg bond lengths remain almost unaltered after hydrogenation.  The slight variations in Ni-Mg bond lengths for $NiMg_2, Ni_2Mg_4$ and $Ni_3Mg_6$ clusters before and after hydrogenation are shown in Figure 2(c).  The bond lengths of $NiMg_2$ match well with previously reported values [ ].  It is observed that $NiMg_2$ cluster can adsorb 9 $H_2$ molecules within the optimum binding energy range.   Similar to the case of 1:1 ratio, the number of adsorbed hydrogen molecules gets doubled and tripled, respectively, with the proportional increment in size of the host $Ni_2Mg_4$ and $Ni_3Mg_6$ clusters from that in $NiMg_2$.  Consequently, $Ni_2Mg_4$ and $Ni_3Mg_6$ are capable of absorbing 18 and 27 $H_2$ molecules and hence the gravimetric density of hydrogen in this alloying ratio is found to be 14.46 wt\%.
\begin{figure}
\centering
\includegraphics[width=0.8\textwidth]{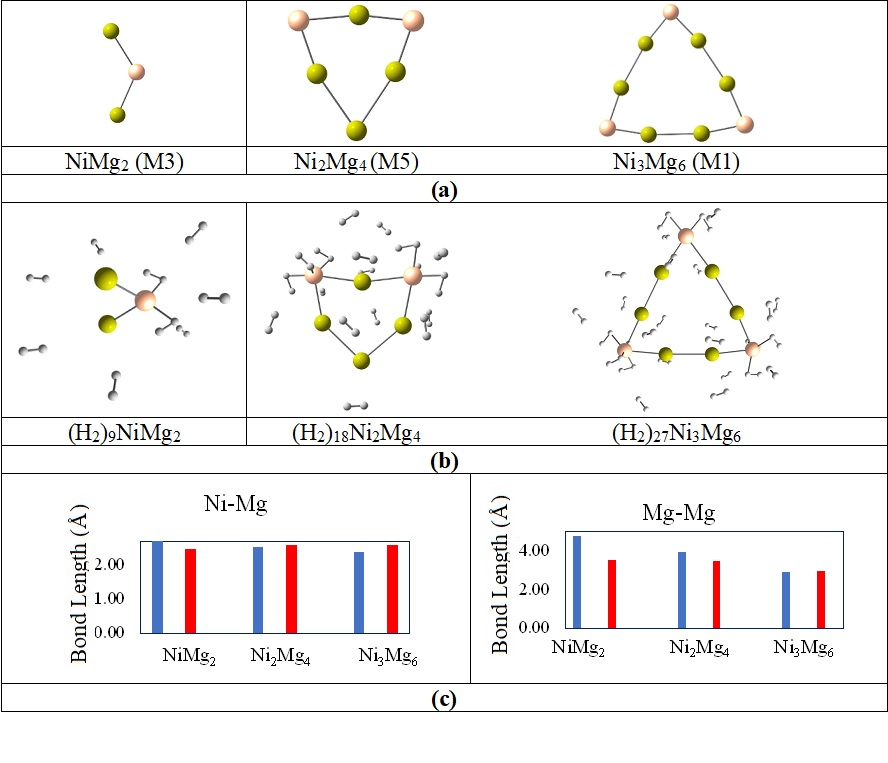}
\caption{\bf Geometries of (a) bare $\bf Ni_nMg_m$ clusters and (b) $\bf (H_2)_xNi_nMg_m$ complexes in 1:2 (Ni:Mg) ratio, (c) Ni-Mg bond lengths of bare and hydrogenated NinMgm clusters. (Colour blue and red are used to represent bare and hydrogenated complexes respectively, M represents spin multiplicity)}\label{Fig2}
\end{figure}
\newpage
\subsubsection {Hydrogen Adsorption on $Ni_nMg_m$ clusters (3n=m, n+m=4k, k=1-4)}
In the 1:3 ratio we considered three compositions viz. $NiMg_3, Ni_2Mg_6$ and $Ni_3Mg_9$. All the structural isomers are shown in figure S3(a).  In this ratio also, Hydrogen adsorption is found to be maximum in the planar isomers with Ni atoms on the peripheral positions.  The structures of bare and hydrogen-adsorbed complexes are shown in figure 3(a) and 3(b).  The variation of Ni-Mg bond lengths of bare and hydrogen absorbed NinMgm clusters are shown in Figure 3(c).  It is observed that the average Ni-Mg bond lengths have not altered much after hydrogenation. The bond length of NiMg3 matches well with already reported literature [53].  Similar to the case of 1:1 and 1:2 (Ni:Mg) ratio, in the $NiMg_3, Ni_2Mg_6$ and $Ni_3Mg_9$ clusters the number of adsorbed hydrogen molecules are found to vary in proportion to the increment in size of the host clusters.  As a result, $NiMg_3, Ni_2Mg_6$ and $Ni_3M_9$ clusters can absorb maximum of 10, 20 and 30 $H_2$ molecules, respectively. The gravimetric density of hydrogen in this Ni:Mg ratio is found to be 13.28 wt\%.  
\newpage
\begin{figure}
\centering
\includegraphics[width=0.8\textwidth]{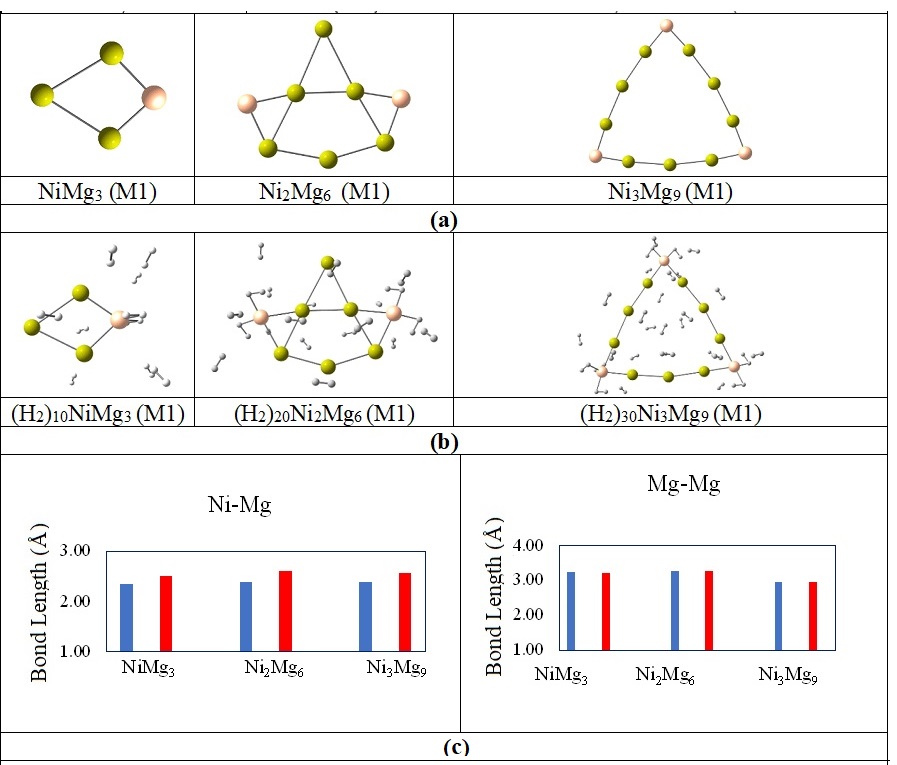}
\caption{\bf Geometries of (a) bare $\bf Ni_nMg_m$ clusters and (b) $\bf (H_2)_xNi_nMg_m$ complexes in 1:3 (Ni:Mg) ratio, (c) Ni-Mg bond lengths of bare and hydrogenated NinMgm clusters. (Colour blue and red are used to represent bare and hydrogenated complexes respectively, M represents spin multiplicity)}\label{Fig3}
\end{figure}
\subsection { Density of States Analysis}
Previous studies have shown that molecular adsorption of hydrogen takes place due to the interaction of the host material with the bonding and anti-bonding orbitals of hydrogen molecule \cite {bib8,bib54}.  To confirm this, we have carried out the partial density of states (PDOS) and the overlap population density of states (OPDOS) calculations of the hydrogenated clusters, which are plotted in Figure 4.  The PDOS and OPDOS are computed using Multiwfn programme \cite {bib55}.  The energy densities are represented by Gaussian distributions with a half width of 1.00 eV.  The OPDOS plots represent the overlapping or linear combination of different atomic orbitals forming the molecular orbitals of the hydrogen adsorbed cluster complexes (Figure 4).  The Fermi level of $(H_2)_9NiMg, (H_2)_{18}Ni_2Mg_2$, and $(H_2)_{27}Ni_3Mg_3$ are observed around -6.70 eV, -6.20 eV and -6.34 eV. The Fermi level of $(H_2)_9NiMg_2$, $(H_2)_{18}Ni_2Mg_4 and (H_2)_{27}Ni_3Mg_6$ complexes are present around -6.23 eV, -4.94 eV and -5.35 eV, whereas that of $(H_2)_{10}NiMg_3, (H_2)_{20}Ni_2Mg_6$ and $(H_2)_{30}Ni_3Mg_9$ are found around -6.20 eV, -5.36 eV and -5.20 eV.  PDOS plots of the Ni and $H_2$ of the hydrogenated $Ni_nMg_m$ clusters are shown in Figure 4.  Sharp larger peaks around -14.00 eV along with smaller peaks at various positions such as ~ -16 eV, -15 eV, -9.00 eV etc. are observed for the PDOS of H(1s) in all the hydrogenated complexes.  For the Ni (3d) PDOS, prominent larger peaks around -9.00 eV with smaller peaks in the range of -16.00 eV to -14.00 eV are found to appear.  It is clear from the same figure that the OPDOS peaks are also present in the energy range as those of H(1s) and Ni (3d) PDOS of the complexes.  Similar interactions are also observed between Mg and $H_2$ molecule (Figure S4).  This confirms the interaction between Ni atoms of NinMgm clusters and $H_2$ molecules.  These interactions are in well agreement with previous studies \cite {bib8, bib50}.
\begin{figure}
\centering
\includegraphics[width=0.8\textwidth]{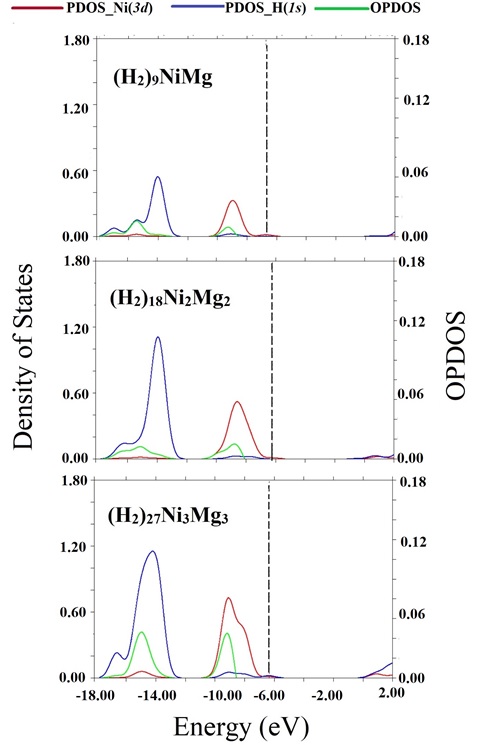}

\end{figure}

\begin{figure}
\centering
\includegraphics[width=0.8\textwidth]{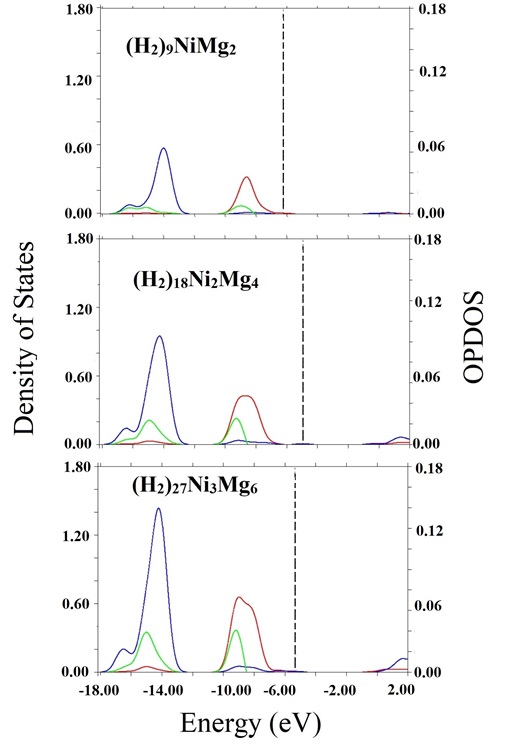}

\end{figure}

\begin{figure}
\centering
\includegraphics[width=0.8\textwidth]{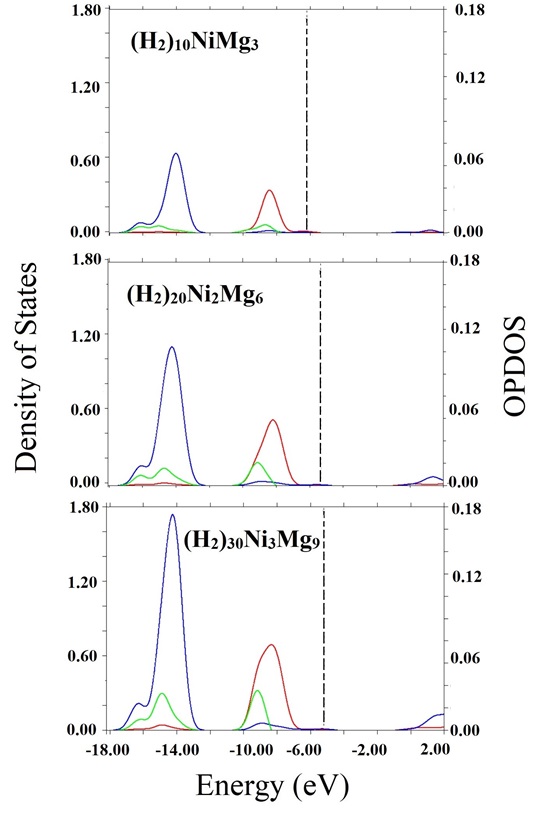}
\caption{\bf PDOS and OPDOS plots of hydrogen adsorbed NinMgm clusters.  The dashed line represents the fermi level.}\label{Fig4}
\end{figure}
\newpage
\subsection {Desorption of Hydrogen}
It is reported that room temperature desorption of hydrogen is possible for adsorption energy range of -0.07 eV to -0.20 eV \cite {bib56,bib57}.  We have already discussed above that the adsorption energy per hydrogen molecule considered in our study is limited to the minimum of -0.10 eV.  Therefore, we can expect the desorption of $H_2$ molecules to take place at room temperature.  To explore the desorption properties at room temperature, Atom Centred Density Matrix Propagation (ADMP) have been performed at 300 K and 1 atm pressure.  ADMP calculations are carried out considering 0.2 fs time step for 300 fs.  The snapshots are provided in a movie format in Figure S5 supporting information.  Figures 5(a)-(c) show the potential energy versus time plot of hydrogen desorption process along with the snapshots at the final studied moment, i.e. at 300 fs of the complexes in Ni:Mg alloying ratios of 1:1, 1:2 and 1:3, respectively.  It is observed that the potential energy changes to a maximum value of 1.36 eV (0.05 au) from the starting point after elapsing approximately fs during the desorption process.  Within the first 50 fs, the complexes carrying Ni:Mg ration of 1:1, $(H_2)_9NiMg$, $(H_2)_{18}Ni_2Mg_2$ and $(H_2)_{27}Ni_3Mg_3$ release 3, 4 and 3 $H_2$ molecules, respectively.  The same complexes are able to release 6, 14 and 15 $H_2$ molecules after 300 fs , respectively.  The complexes having 1:2 (Ni:Mg) ratio, $(H_2)_9NiMg_2$, $(H_2)_{18}Ni_2Mg_4$ and $(H_2)_{27}Ni_3Mg_6$ release 3, 3 and 1 $H_2$ molecules, respectively in the first 50 fs, which increase to 7, 13 and 16 H2 molecules after 300 fs.  The complexes studied in the 1:3 (Ni:Mg) ratio, $(H_2)_{10}NiMg_3$, $(H_2)_{20}Ni_2Mg_6$ and $(H_2)_{30}Ni_3Mg_9$ release 3 $H_2$ molecules each in 50 fs and they releases 8, 14 and 19 $H_2$ molecules, respectively after 300 fs.  Therefore, it can be confirmed that for all the complexes 
\begin{figure}
\centering
\includegraphics[width=0.8\textwidth]{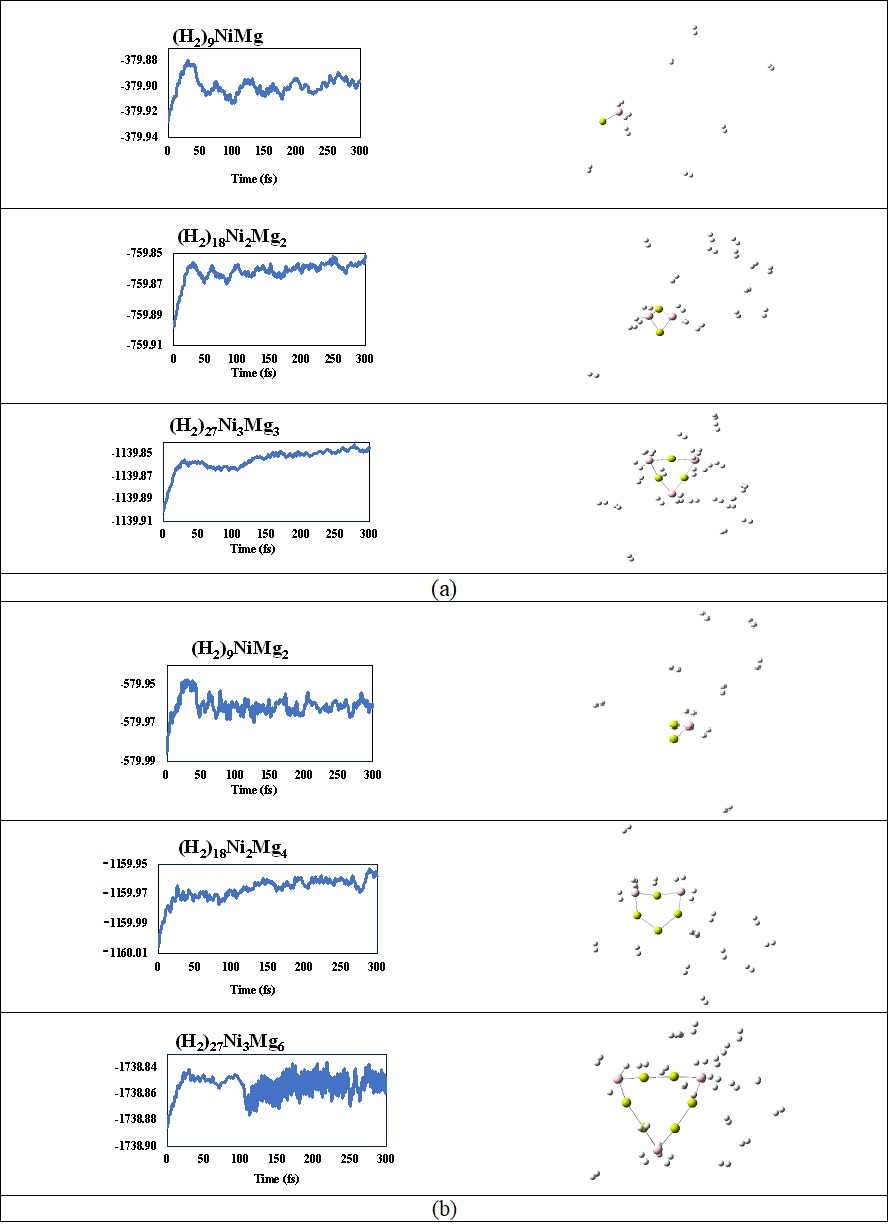}
\end{figure}
\newpage
\begin{figure}
\centering
\includegraphics[width=0.8\textwidth]{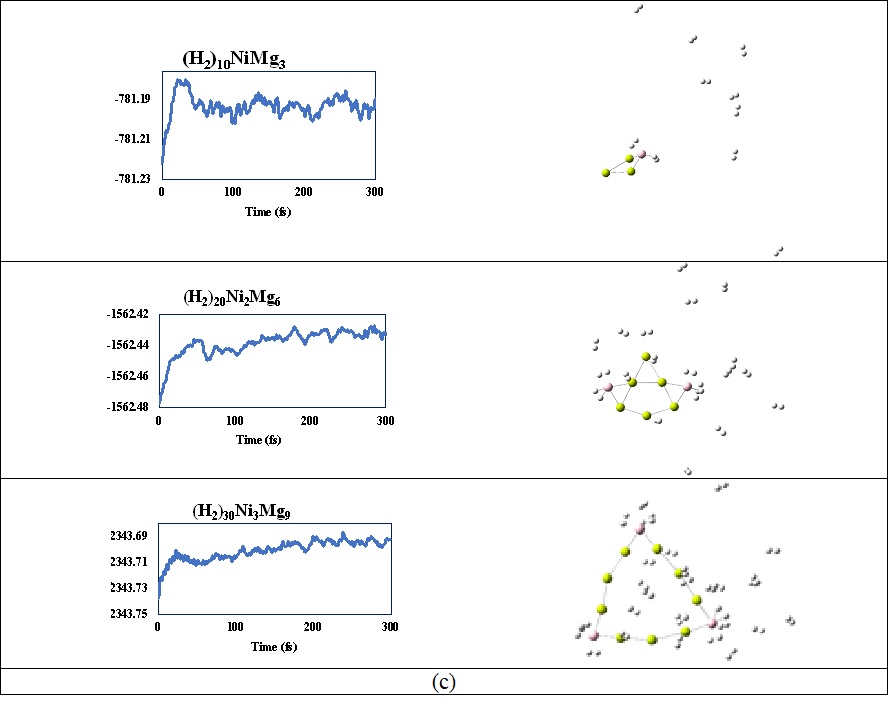}
\caption{\bf ADMP potential energy versus time plots and snapshots at 300 fs of hydrogen dissociation on $\bf Ni_nMg_m$ clusters at (a) 1:1 (b)1:2 and (c) 1:3 alloying ratios.}\label{Fig5}
\end{figure}
\newpage
\section {Conclusion}
Symmetry unrestricted full geometry optimizations have been carried out in different spin multiplicities for a series of Ni doped small Mg clusters, $Ni_nMg_m$ maintaining the Ni:Mg ratio to be 1:1, 1:2 and 1:3.  For each of these ratios, three different cluster sizes have been considered in our study for studying the nature of adsorption of multiple hydrogen molecules.  It is observed that the planar structures with sufficiently exposed Ni atoms are only able to adsorb the maximum number of $H_2$ molecules.  However, these cluster isomers are other than the lowest energy isomers except for NiMg dimer in 1:1 ratio.   Except for $NiMg_2$ cluster the geometries of these clusters remain almost unaltered after hydrogenation as shown by negligible changes in their bond lengths, which confirm the absence of chemisorption of any $H_2$ molecule on the clusters.  The calculated binding energy (BE) values fall in the range between physisorption and chemisorption processes.  The minimum cut off value on BE has been chosen to be 0.1 eV to find the maximum number of adsorbed $H_2$ molecules maintaining the desired range of $0.1 eV \geq BE \geq0.8 eV$ for suitable hydrogen storage materials.  Our computed results show that $Ni_kMg_k$ (Ni:Mg=1:1) and $Ni_kMg_{2k}$ (Ni:Mg=1:2) clusters adsorb 9, 18 and 17 $H_2$ molecules for k=1, 2 and 3, respectively.  Again, $Ni_kMg_{3k}$ (Ni:Mg=1:3) clusters can adsorb 10 (k=1), 20 (k=2) and 30 (k=3) $H_2$ molecules.  The important feature of this observation is that for a particular Ni:Mg alloying ratio, the number of adsorbed hydrogen molecules grows in a multiplicative manner as the clusters size increases.  As a result, hydrogen gravimetric density does not vary with cluster size provided the alloying ratio is fixed.  The gravimetric densities for clusters in 1:1, 1:2 and 1:3 ratios of Ni:Mg are found to be 17.94 wt\%, 14.46 wt\% and 13.28 wt\%, respectively, which are fairly above the targeted value of 6.5 wt\% set by Department of Energy (DOE), US.  Therefore, we can conclude that linear and planar-like structures of $Ni_nMg_m$ clusters bearing exposed Ni atoms are able to adsorb significantly larger number of hydrogen molecules with 1:1 alloying ratio being the most effective one.  Further, molecular dynamics simulations have shown room temperature desorption of almost all the $H_2$ molecules from the studied clusters at standard pressure.  These results favour possible reversible use of such small NinMgm clusters under ambient thermodynamic conditions.

\paragraph {Acknowledgement}

BB thanks University Grants Commission, India for research fellowship.


\newpage
\begin{thebibliography}{9}
\bibitem{bib1}
 https://www.energy.gov/eere/fuelcells/hydrogen-storage

\bibitem{bib2} S. K. Bhatia, A. L. Myers, Optimum conditions for adsorptive storage, Langmuir. 22 (2006)1688 – 1700.
 
\bibitem{bib3} R.C. Lochana, M. Head-Gordon, Computational studies of molecular hydrogen binding affinities: the role of dispersion forces, electrostatics and orbital interactions, Phys. Chem. Chem. Phys. 8 (2006)1357 – 1370.

\bibitem{bib4} N. Yuksel, A. Kose, M. F. Fellah, A density functional theory study of molecular hydrogen adsorption on Mg site in OFF type zeolite cluster, Int. J. Hydrogen Energy. 45 (2020) 34983 – 34992.

\bibitem{bib5} K. Gopalsamy, V. Subramanian, Hydrogen storage capacity of alkali and alkaline earth metal ions doped carbon based materials: A DFT study, Int. J. Hydrogen Energy. 39 (2014) 2549 – 2559.

\bibitem{bib6} A. Kumar, N. Vyas, A. K. Ojha, Hydrogen storage in magnesium decorated boron clusters ($Mg_2Bn, n=4-14$):A density functional theory study, Int. J. Hydrogen Energy. 45 (2020) 12961 – 12971.

\bibitem{bib7} K. Srinivasu, S. K. Ghosh, R. Das, S. Giri, P. K. Chattaraj, Theoretical investigation of hydrogen adsorption in all-metal aromatic clusters, RSC Advances. 2 (2012) 2914 – 2922.

\bibitem{bib8} B. Boruah, B. Kalita, Exploring enhanced hydrogen adsorption on Ti doped Al nanoclusters: A DFT study, Chem. Phys. 518 (2019) 123-133.

\bibitem{bib9} A. Jaiswal, Rakesh K. Sahoo, S. S. Ray and S. Sahu, Alkali metal decorated silicon clusters ($Si_nMn$, n=6, 10: M=Li,Na) as potential hydrogen storage materials: A DFT study, Int J. Hydrogen Energy. 47(3) (2022), 1775-1789.

\bibitem{bib10} R. K. Sahoo, B Chakraborty, S. Sahu, Reversible hydrogen storage on alkali metal (Li and Na) decorated $C_{20}$ fullerene: A density functional study, Int. J. Hydrogen Energy. 46 (2021) 40251– 40261.  

\bibitem{bib11} J. Du, X. Sun, L. Zhang, C. Zhang, G. Jiang, Hydrogen storage of $Li_4$ and $B_{36}$ cluster, Sc. Reports 8, (2018) 1940-1 –1940-7.  

\bibitem{bib12} X. B. Xie et al., First principles studies in Mg-based hydrogen storage materials: A review, Energy. 211 (2020) 118959.

\bibitem{bib13} X. Wang, L. Andrews, Infrared spectra of magnesium hydride molecules, complexes and solid magnesium hydride, J. Phys. Chem. A. 108(2004) 11511 – 11520.

\bibitem{bib14}B. Li  et al.,  Mg-based metastable nano alloys for hydrogen storage, Int. J. Hydrogen Energy.  44 (2019) 6007 – 6018.

\bibitem{bib15} V. A. Yartys et. al, Magnesium based materials for hydrogen storage: Past, present and future, Int. J. Hydrogen Energy. 44 (2019) 7809 – 7859.

\bibitem{bib16} D. He, Y. Wang, C. Wu, Q. Li, W. Ding, C. Sun, Enhanced hydrogen desorption properties of magnesium hydride by coupling non-metal doping nano-confinement, Appl. Phys. Lett. 107 (2015) 243907-1 – 243907-5.

\bibitem{bib17} Z. Sun, X. Lu, F. M. Nyahuma, N. Yan, J. Xiao, S. Su, L. Zhang, Enhancing hydrogen storage properties of $MgH_2$ by transition metals and carbon materials: A brief review, Front. Chem. 8 (2020) 552-1-552 – 552-14. 

\bibitem{bib18} R. W. P. Wagemans, J. H. van Lanthe, P.E. de Jongh, A. J. van Dillen, K. P. de Jong, Hydrogen storage in magnesium clusters: Quantum chemical study, J. Am. Chem. Soc. 127 (2005) 16675 – 16680.

\bibitem{bib19} I. P. Jain, C. Lal, A. Jain, Hydrogen storage in Mg: A most promising material, Int. J. Hydrogen Energy. 35 (2010) 5133 – 5144.

\bibitem{bib20} J. Liu, J. Tyrrell, L. Cheng, Q. Ge, First-principles studies on hydrogen desorption mechanism of $Mg_nH_{2n}$ (n=3,4), J. Phys. Chem. C. 117 (2013) 8099 – 8104. 

\bibitem{bib21} W. Ma, C. Jing, First-principles study on hydrogen storage in Al-, Ca-, Mn- doped MgNi clusters, Int. J. Mod. Phys. B 31 (2017) 1730002-1 – 1730002-10.

\bibitem{bib22}X. Zhang, Y. Liu, Z. Ren, et al. Realizing 6.7 wt\% reversible storage of hydrogen at ambient temperature with non-confined ultrafine magnesium hydrides, Energy Environ. Sci. 14 (2021) 2302 – 2313.

\bibitem{bib23} S. Hao, D. S. Sholl, Selection of dopants to enhance hydrogen diffusion rates in $MgH_2$ and $NaMgH_3$. Appl. Phys. Lett., 94 (2009) 17190-1– 17190-3.

\bibitem{bib24}W Li, C. Li, H. Ma, J. Chen, Magnesium nanowires: Enhanced kinetics for hydrogen adsorption and desorption, J. Am. Chem. Soc. 129 (2007) 6710 – 6711. 

\bibitem{bib25} E. N. Kaukaras, A. D. Zdetsis, M. M. Sigalas, Ab initio study of magnesium and magnesium hydride nanoclusters and nanocrystals: Examining optimal structures and compositions of efficient hydrogen storage, J. Am. Chem. Soc. 134 (2012) 15914 – 15922.

\bibitem{bib26}N. S. Norberg, T. S. Arthur, S. J. Fredrick, A. L. Prieto, Size dependent hydrogen storage properties of Mg nanocrystals prepared from solution, J. Am. Chem. Soc. 133 (2011) 10679 – 10681.

\bibitem{bib27} F. Cheng, Z. Tao, J. Liang, J. Chen, Efficient hydrogen storage with the combination of lightweight Mg/MgH2 and nanostructures, Chem. Commun. 48 (2012) 7334 – 7343.

\bibitem{bib28} K. J. Jeon, H. R. Moon, A. M. Ruminski, B. Jiang, C. Kisielowski, R. Bardhan, J. J. Urban, Air-stable magnesium nanocomposites provide rapid and high-capacity hydrogen storage without using heavy metal catalysts. Nat. Mater. 10 (2011) 286 – 290.

\bibitem{bib29} D. Shen, C. P. Kong, R. Jia, P. Fu, H. X. Zhang, Investigation of properties of $Mg_n$ clusters and their hydrogen storage mechanism: A study based on DFT and Global minimum optimisation method, J. Phys. Chem. A. 119 (2015) 3636-3643.

\bibitem{bib30} P. Banerjee, K. R. S. Chandrakumar, G. P. Das, Exploring adsorption and desorption characteristics of molecular hydrogen on neutral and charged Mg nanoclusters: A first principles study, Chem. Phys. 469-470 (2016) 123 – 131.

\bibitem{bib31} A. K. Srivastava, N. Misra, Ab initio investigations on planar $(MgO)_n$ clusters (n=1-5) and their hydrogen adsorption behaviour, Molecular Simulations. 42(2016):208 – 214.

\bibitem{bib32} M. El Khatabi, M. Bhihi, S. Naji, H. Labrim, A. Benyoussef, A. El kenz, M. loulidi, Study of doping effects with 3d and 4d-transition metals on the hydrogen storage properties of $MgH_2$, Int. J. Hydrogen Energy. 41(2016):4712 – 4718.

\bibitem{bib33} Y. Wnag, S. Lu, Z. Zhou, G. jin, Z. Lan, Effect of transition metal on hydrogen storage properties of Mg-Al alloy, J. Mater. Sci. 52 (2017) 2392 – 2399.  

\bibitem{bib34} R. Trivedi, D. Bandyopadhyay, Study of adsorption and dissociation pathway of $H_2$ molecule on $Mg_nRh$ (n=1-10) clusters: A first principle investigation, Int. J. Hydrogen Energy. 41 (2016) 20113 – 20121.

\bibitem{bib35} M. Pozzo, D. Alfe, A. Amieiro, S. French, A. Pratt, Hydrogen dissociation and diffusion on Ni- ,Ti- doped Mg(001) surfaces, J. Chem. Phys. 128 (2008) 094703-1 – 094703-1-11.

\bibitem{bib36} R. Trivedi, D. Bandyopadhyay, Hydrogen adsorption in small size $Mg_nCo$ clusters: A density functional study, Int. J. Hydrogen Energy. 40 (2015;) 12727 – 12735.

\bibitem{bib37} X. Xie, M. Chen, M. Hu, B. Wang, R. Yu, T. Liu, Recent advances in Mg-based hydrogen storage materials with multiple catalysts, Int. J. Hydrogen Energy. 44 (2019) 10694 – 10712.

\bibitem{bib38} X. Ma, S. Liu, S. Huang, Hydrogen adsorption and dissociation on the TM-doped (TM=Ti, Nb) $Mg_{55}$ nanoclusters: A DFT study, Int. J. Hydrogen Energy. 42 (2017) 24797 – 24810.

\bibitem{bib39} Y. Wu, Y. Meng, L. Ma, J. Zhao, J. Tang, H. Chen, How does Ti-doping effect hydrogen storage properties of $MgH_2$ at nanosize, Russ. J. Phys. Chem. A. 95(2021) 1424 – 1431. 

\bibitem{bib40} S. S. Samantaray, P. Anees, V. B. Parambath et al., Graphene supported MgNi alloy composites as room temperature hydrogen storage material 	- experiments and theoretical insights, Acta Material. 215 (2021) 117040-1 – 117040-3.

\bibitem{bib41} A. P. Maltsev, O.P. Charkin, Theoretical modelling of stepwise addition of $H_2$ molecules to magnesium clusters $Mg_{18}$ and $Mg_{17}M$, Russ. J.  Inorg. Chem. 65 (2020)185 – 192.

\bibitem{bib42} A. P. Maltsev, O.P. Charkin, Theoretical modelling of stepwise addition of $H_2$ molecules to $Mg_{17}L$ magnesium clusters doped with 3d metals, Russ. J.  Inorg. Chem. 65 (2020) 1204 –1211.

\bibitem{bib43}  O.P. Charkin, A. P. Maltsev, Density functional theory modelling of reactions of addition of H2 molecules to magnesium clusters \%Mg17M doped with atoms M of transition 3d elements, J. Phys. Chem. A. 125 (2021) 2308 – 2315.

\bibitem{bib44}  O. P. Charkin, A. P. Maltsev, Theoretical modelling of exo- and endohedral hydrogenation reactions of the doped Mg cluster $Mg_{17}Ni$, Russ. J.  Inorg. Chem. 66 (2021) 1860 – 1867. 

\bibitem{bib45} Gaussian 09, Revision D.01, Frisch MJ, Trucks GW, Schlegel HB, Scuseria GE, Robb MA, Cheeseman JR, Scalmani G, Barone V, Mennucci B, Petersson GA, et al. Gaussian, Inc., Wallingford CT, 2009.

\bibitem{bib46} B. Boruah, B. Kalita, Role of transition metal doping in determining the electronic structure and properties of small magnesium clusters: a DFT-based comparison of neutral and cationic states, J. Nanopart. Res. 22 (2020) 370-1 – 370-19.

\bibitem{bib47} J. D. Chai, M. H. Gordon, Long-range corrected hybrid density functionals with damped atom-atom dispersion corrections, Phys. Chem. Chem. Phys. 10 (2008) 6615 – 6620.

\bibitem{bib48} Y. Minekov, A. Singstad, G. Occhipinti, V. R. Jensen, The accuracy of DFT-optimised geometries of functional transition metal compounds: a validation study of catalysts for olefin metathesis and other reactions in the homogeneous phase, Dalton Trans. 41 (2012) 5526 – 5541.

\bibitem{bib49} J. Lv, Y. wang, L. Zhu, Y. Ma, Particle-swarm structure prediction on clusters, J. Chem. Phys. 137 (2012) 084104-1 – 084104-8

\bibitem{bib50} arXiv:2112.09855v1 [cond-mat.mtrl-sci] 

\bibitem {bib51} L. Jing, L. Xiao-Yong, Z. Zheng-He, S. Yong, Density functional theory study of $Mg_nNi_2$ (n=1-6) clusters, Chin. Phys. B. 21(3) (2012) 033101-1 – 033101-7. 

\bibitem {bib52} C. Xue-Feng, Z. Yan, Q. Kai-Tian, L. Bing et al., Density functional theory study on Ni-doped $Mg_nNi$ (n=1-7) clusters, Chin. Phys. B. 19(3) (2010) 033601-1 – 033601-5.

\bibitem {bib53} Deepika, K. Raj R, T. J. D. Kumar, R. Kumar, Sequential desorption energy of hydrogen from Ni clusters, AIP Conf. Proc. 1665 (2015), 080076-1 – 080076-3.

\bibitem {bib54} G. J. Kubas, Metal-dihydrogen and $\sigma$-bond coordination: the consummate extension of the Dewar-Chatt-Duncanson model for metal olefin $\pi$-boding, J. Organ. Chem. 635 (2001) 36 – 68. 

\bibitem {bib55} Lu T, Chen F (2012) Multifwfn: A Multifunctional Wavefunction Analyzer. J. Comput. Chem. 33:580 –592.

\bibitem {bib56} R. Y. Sathe, S. Kumar, T. J. D. Kumar, An ab initio study of reversible hydrogen adsorption in metal decorated $\gamma$-graphyne, J. Appl. Phys. 126(7) (2019) 174301-1–174301-10.

\bibitem {bib57} P. Banerjee, B. Pathak, R. Ahuja, G. P. Das, First principles design of Li functionalized hydrogenated h-BN nanosheet for hydrogen storage, Int. J. Hydrogen Energy. 41(32) (2016) 14437-14446.

\end{thebibliography}
\end{document}